\documentclass[twocolumn,usenatbib]{mnras}

\usepackage{graphicx}
\usepackage{amsmath}
\usepackage{comment}
\usepackage{hyperref}

\def\pmb#1{\setbox0=\hbox{#1}%
    \kern-.025em\copy0\kern-\wd0
    \kern.05em\copy0\kern-\wd0
    \kern-.025em\raise.0433em\box0}

\def\ltsima{$\; \buildrel < \over \sim \;$}
\def\gtsima{$\; \buildrel > \over \sim \;$}
\def\simlt{\lower.5ex\hbox{\ltsima}}
\def\simgt{\lower.5ex\hbox{\gtsima}}
\def\p2Y{\;_2Y}
\def\m2Y{\;_{-2}Y}

\def\mk2{\mu {\rm K}^2}
\def\Planck{\it Planck \rm}

\def\Planckns{\it Planck\rm}
\def\LCDM{$\Lambda{\rm CDM}$ }
\def\LCDMns{$\Lambda{\rm CDM}$}

\newcommand{\Mpc}{\text{Mpc}}

\def\pmb#1{\setbox0=\hbox{#1}%
     \kern-.025em\copy0\kern-\wd0
     \kern.05em\copy0\kern-\wd0
     \kern-.025em\raise.0433em\box0}

\begin{document}

\title[$S_8$ tension]{A non-linear solution to the $S_8$ tension?}

\author[Amon \& Efstathiou]{Alexandra Amon$^{1}$\thanks{E-mail: alexandra.amon@ast.cam.ac.uk}, George Efstathiou$^{1}$\thanks{E-mail: gpe@ast.cam.ac.uk} \\
${1}$ Kavli Institute for Cosmology Cambridge, 
Madingley Road, Cambridge, CB3 OHA.}

\maketitle

\begin{abstract} 
  Weak galaxy lensing surveys have consistently reported a lower amplitude for the matter fluctuation spectrum, as measured by the $S_8$ parameter, than expected in the \LCDM cosmology favoured by \textit{Planck}.
  However, the expansion history  follows the predictions of the \textit{Planck} \LCDM cosmology to high accuracy,  as do measurements of lensing of the cosmic microwave background anisotropies. 
Redshift space distortion measurements also appear to be consistent with \textit{Planck} \LCDMns.  
  In this paper, we argue that  these observations can be reconciled with the \textit{Planck} \LCDM cosmology if the matter power spectrum is suppressed more strongly on non-linear scales than assumed in analyses of weak galaxy lensing.  We demonstrate this point by fitting a one-parameter model, characterising a  suppression of the non-linear power spectrum, 
  to the KiDS-1000 weak lensing measurements. Such a suppression could be attributed to  new properties of the dark matter that affect non-linear scales, or to a response of the matter fluctuations to baryonic feedback processes that are stronger than expected from recent cosmological simulations. Our proposed explanation can be tested using measurements
  of the amplitude of the matter fluctuation spectrum on linear scales, in particular via high precision redshift space distortion measurements
  from forthcoming galaxy and quasar redshift surveys.

\end{abstract}


\section{Introduction}\label{sec:intro}

The standard \LCDM cosmological model  provides a remarkably good fit to a number of observations, including anisotropies of the cosmic microwave background  \citep[CMB; e.g.][]{Bennett:2013, Params:2018, Aiola:2020,   Dutcher:2021}, baryon acoustic oscillations \citep[BAO; e.g.][] {Alam:2017, Blomqvist:2019, deSainteAgathe:2019, Alam:2021} and the magnitude-redshift relation of Type Ia supernovae \citep[e.g.][]{Betoule:2014, Scolnic:2017, Brout:2021}. Despite these successes, there is some evidence of `tensions' with other astrophysical data at varying degrees of statistical significance. 

At the level of the background expansion history, some late time measurements of the Hubble constant disagree with the \LCDM value inferred from the CMB \citep[e.g.][]{Riess:2016, Riess:2019, Riess:2021, Wong:2020} though others do not \citep{Freedman:2019, Freedman:2020}. This problem has become known as the `Hubble tension' \cite[for recent reviews see][]{Freedman:2021, Shah:2021, Snowmass:2022}.

At the level of perturbations, weak lensing surveys have  reported
measurements of the  amplitude of the fluctuation spectrum  
 \citep[e.g.][]{Heymans:2013,  Hikage:2019,  Hamana:2020, Asgari:2021, Amon:2021, Secco:2022}. 
 Specifically, cosmic shear surveys tightly constrain the parameter combination\footnote{Where $\Omega_{\rm m}$ is the present day matter density in units of the critical density, $\sigma_8$ is the root mean square linear  amplitude of the matter fluctuation spectrum in spheres of radius $8 h^{-1} {\rm Mpc}$ extrapolated to the present day, and $h$ is the value of the Hubble constant $H_0$ in units of $100 \ {\rm km}{\rm s}^{-1}{\rm Mpc}^{-1}$.} $S_8 = \sigma_8 (\Omega_{\rm m}/0.3)^{0.5}$,  consistently finding values that are lower than that  expected according to  the \textit{Planck} best fit \LCDM\ cosmology.
 This discrepancy has become known as the `$S_8$ tension'. 
Most recently,  two large cosmic shear surveys have reported new constraints. Assuming a spatially flat \LCDM cosmology, the  Kilo-Degree Survey (KiDS)-1000 gives\footnote{We quote results derived from the COSEBI (complete orthogonal sets of E/B integrals, see \cite{Schneider:2010}) statistics.}:
\begin{equation}
S_8 = \left \{ \begin{array}{ll}   0.759^{+0.024}_{-0.021}, &  \mbox{KiDS--1000  cosmic shear}, \\
                                   0.766^{+0.020}_{-0.014}, &  \mbox{KiDS--1000  {3$\times$2}pt, } \end{array}\right.  \label{equ:KiDS-1}
\end{equation}
see \cite{Asgari:2021} for the shear-shear analysis (hereafter KiDS21) and  \cite{Heymans:2021} for a $3\times2{\rm pt}$ analysis combining  shear-shear, galaxy-galaxy lensing and galaxy-galaxy two-point statistics. The Dark Energy Survey (DES) Year 3 analysis gives 
\begin{equation}
S_8 = \left \{ \begin{array}{ll}   0.772^{+0.018}_{-0.017}, &  \mbox{DES Y3  cosmic shear}, \\
                                   0.779^{+0.014}_{-0.015},  &  \mbox{DES Y3  {3$\times$2}pt } \end{array}\right.   \label{equ:DES-Y3}
\end{equation}
where we have quoted the `\LCDM optimised' results for the shear-shear analysis \citep{Amon:2021, Secco:2022}, hereafter DES22, and for the 3x2pt function analysis \citep{DES:2021}. The results quoted above are derived using different analysis choices (angular scale cuts, intrinsic alignment model, etc.) and differences in assumptions concerning neutrino masses\footnote{The \Planck and KiDS analyses assume a normal hierarchy with the  heaviest neutrino mass  fixed to $0.06\ {\rm eV}$, while DES allows the neutrino mass to vary, which has little impact on their $S_8$ constraints \cite[][Fig.~28]{DES:2021}.}.

For \textit{Planck},  we adopt  the \LCDM parameters
reported in \cite{EfstathiouGratton:2021} (hereafter EG21):
\begin{subequations}
\begin{eqnarray}
       S_8 & =&  0.828 \pm 0.016, \qquad {\rm TTTEEE},  \label{equ:S8b} \\
       S_8 & =&  0.829 \pm 0.012, \qquad {\rm TTTEEE+Plens},  \label{equ:S8c}
\end{eqnarray}
\end{subequations}
where TTTEEE denotes the high multipole likelihood constructed by combining the temperature power spectra (TT), temperature-polarization E-mode cross-spectra (TE) and polarization E-mode power spectra (EE). Each of these likelihoods is combined with low multipole ($\ell \le 29$ ) TT and EE likelihoods described in \cite{Params:2018}.
`Plens' in Eq. (\ref{equ:S8c}) denotes the addition of the \textit{Planck} CMB lensing likelihood \citep{Plensing:2020}.  

The KiDS-1000 cosmic shear measurements of $S_8$ are about 9\% lower than the \Planck value  of Eq. (\ref{equ:S8c}), suggesting a discrepancy at the $\sim 2.4-2.7 \sigma$ level\footnote{We quote $\Delta S_8/\sqrt{(\sigma^{\rm Planck}_{S_8})^2 + (\sigma_{S_8}^{\rm lensing})^2}$.} depending on which of the KiDS-1000 and \Planck measurements are used in the comparison.  This is consistent with the conclusions of the KiDS team based on more complex  tension metrics \citep{Asgari:2021, Heymans:2021}. The DES Y3  measurements are about $7\%$ lower than the \Planck value, suggesting a discrepancy  at about the $\sim 2.3-2.6 \sigma$ level\footnote{The DES team exclude \Planck lensing when comparing their results with \textit{Planck}. As can be seen from Eqs (\ref{equ:S8b})-(\ref{equ:S8c}) adding \Planck lensing  reduces the error on $S_8$, increasing the significance of the discrepancy with the DES Y3 results.}. To simplify the analysis in this paper, we focus on the cosmic shear measurement since it is the dominant contribution in the $3\times2{\rm pt}$ $S_8$ constraint and therefore the driver of the $S_8$ tension.

While neither of the lensing surveys taken in isolation offers decisive evidence for a discrepancy with the \textit{Planck} \LCDM cosmology, both surveys find low values of $S_8$ in agreement with earlier work. It therefore seems unlikely  that the $S_8$ tension is simply a statistical fluctuation. However, we note that it is naive to crudely combine the shear estimates in Eqs (\ref{equ:KiDS-1}) and (\ref{equ:DES-Y3}), as these analyses use different modelling frameworks, cosmological priors, angular ranges
and various other analysis choices such as to the modelling of intrinsic alignments (IA)\footnote{A combined shear-shear analysis of DES Y3 and KiDS-1000, including a detailed analysis of the differences in methodology and analysis choices, is currently in preparation by the KiDS and DES collaborations.}.

In the last few years there have been significant advances in the calibration of the lensing data \citep[see][and references therein]{Asgari:2020, Amon:2021}. Improved methods of calibrating photometric redshifts  \citep{Hildebrandt:2021, Myles:2021} show
that errors in the redshift distributions of the source galaxies are
unlikely to account for a $\sim 7\%-9\%$ discrepancy in value of $S_8$. It also seems implausible that systematic errors in the shear measurements could explain  a discrepancy of this size \citep{Mandelbaum:2018, Kannawadi:2019, MacCrann:2022}. Similarly, the modelling of IA may affect the value of $S_8$ at the one or two percent level (given the small amplitudes of the alignment corrections found by KiDS and DES) but it seems unlikely that IA are responsible for a $\sim 7\%$ discrepancy. 

What is the most likely explanation of this $S_8$ tension? Does it
require a radical departure from the \LCDM paradigm?  We explore both
of these questions in this paper. This  paper is structured as
follows. Section 2 discuss some preliminaries including comparisons with
other types of data which serve as pointers towards an explanation. The
main results of this paper are contained in Section 3, which explores
the sensitivity of weak lensing results to the modelling of the
 matter power spectrum on non-linear scales. We discuss the implications of our findings in Section 4 and 
summarize our conclusions  in Section 5.

\section{Preliminaries}
\label{sec:preliminaries}

\subsection{The $S_8$ tension}

\begin{figure}
	\centering
	\includegraphics[width=\columnwidth, angle=0]{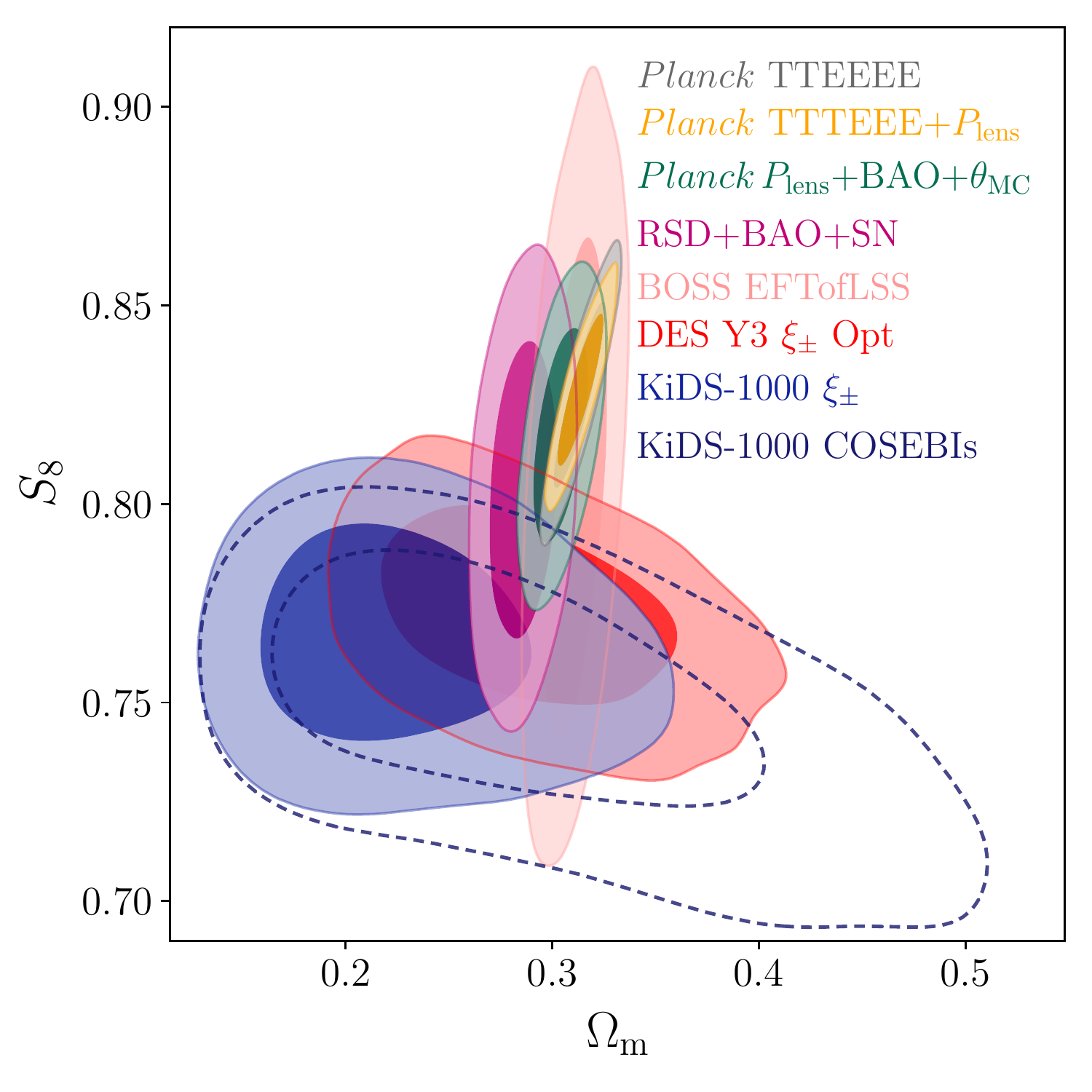} 
	\caption{68\% and 95\% constraints in the $S_8-\Omega_{\rm m}$ plane for various data. The blue and navy (dashed) show the constraints from the KiDS $\xi_{\pm}$ and COSEBI statistics as analyzed by KiDS21, while the red shows that from the DES Y3 \LCDM optimised $\xi_{\pm}$ analysis. The yellow and grey contours show constraints from \Planck TTTEEE with and without the addition of the \Planck CMB lensing likelihood (Plens). The peach contours labelled EFTofLSS represent constraints from the BOSS power spectrum and bispectrum effective field theory analysis of \citet{DAmico:2022}. The magenta contours show constraints from redshift space distortions (RSD) combined with BAO and SN measurements as described in the text. The green contours show the constraint from the \Planck lensing likelihood combined with BAO together with conservative priors on the acoustic peak location parameter $\theta_{\rm MC}$ and other cosmological parameters \citep{Plensing:2020}.}

	\label{fig:S8}
	
\end{figure}

In this paper, we focus exclusively on the KiDS-1000 shear-shear
results, adopting their fiducial analysis pipeline and modelling choices\footnote{KiDS Cosmology
Analysis Pipeline: \url{https://github.com/KiDS-WL/kcap}}.
Fig.~\ref{fig:S8} shows the KiDS21 constraints
in the $S_8$-$\Omega_{\rm m}$ plane from the publicly released MCMC
chains\footnote{\url{http://kids.strw.leidenuniv.nl/sciencedata.php}}.
\cite{Asgari:2021} present results for
three two-point statistics: the usual shear two-point correlation
functions $\xi_{\pm}$, COSEBIs (which KiDS21 adopt as their default for parameter analysis) and angular power spectra. The COSEBIs and power spectra are estimated by integrating over the correlation functions, thus any model that fits $\xi_{\pm}$ over the full angular range probed by KiDS should be consistent with the COSEBIs and power spectra.  Fig.~\ref{fig:S8} also shows constraints from \Planckns. The red contours show the constraints from the DES Y3 shear-shear analysis
\citep{Amon:2022, Secco:2022}. As noted above, $S_8$ from DES Y3 is slightly higher but statistically consistent with the values determined from the KiDS analysis.

The $S_8$ tension between \textit{Planck} and cosmic shear summarized in
Sect.~\ref{sec:intro} is clearly evident and is particularly
acute if one compares \Planck TTTEEE+Plens to the KiDS $\xi_{\pm}$
results. One can also see that the KiDS $\xi_{\pm}$ contours, whilst
overlapping with the COSEBI contours, are shifted to lower values of
$\Omega_{\rm m}$. The $\xi_{\pm}$ and the COSEBI analyses differ primarily in the range of spatial scales sampled by the statistics, with the $\xi_{\pm}$ contours more sensitive to small scales \citep{Asgari:2021}. 
We will focus on the KiDS $\xi_{\pm}$ measurements in this paper, though we discuss the COSEBI measurements in Appendix~\ref{sec:cosebis}.

\subsection{Constraints from redshift space distortions }

\begin{figure}
	\centering
     \includegraphics[width=\columnwidth]{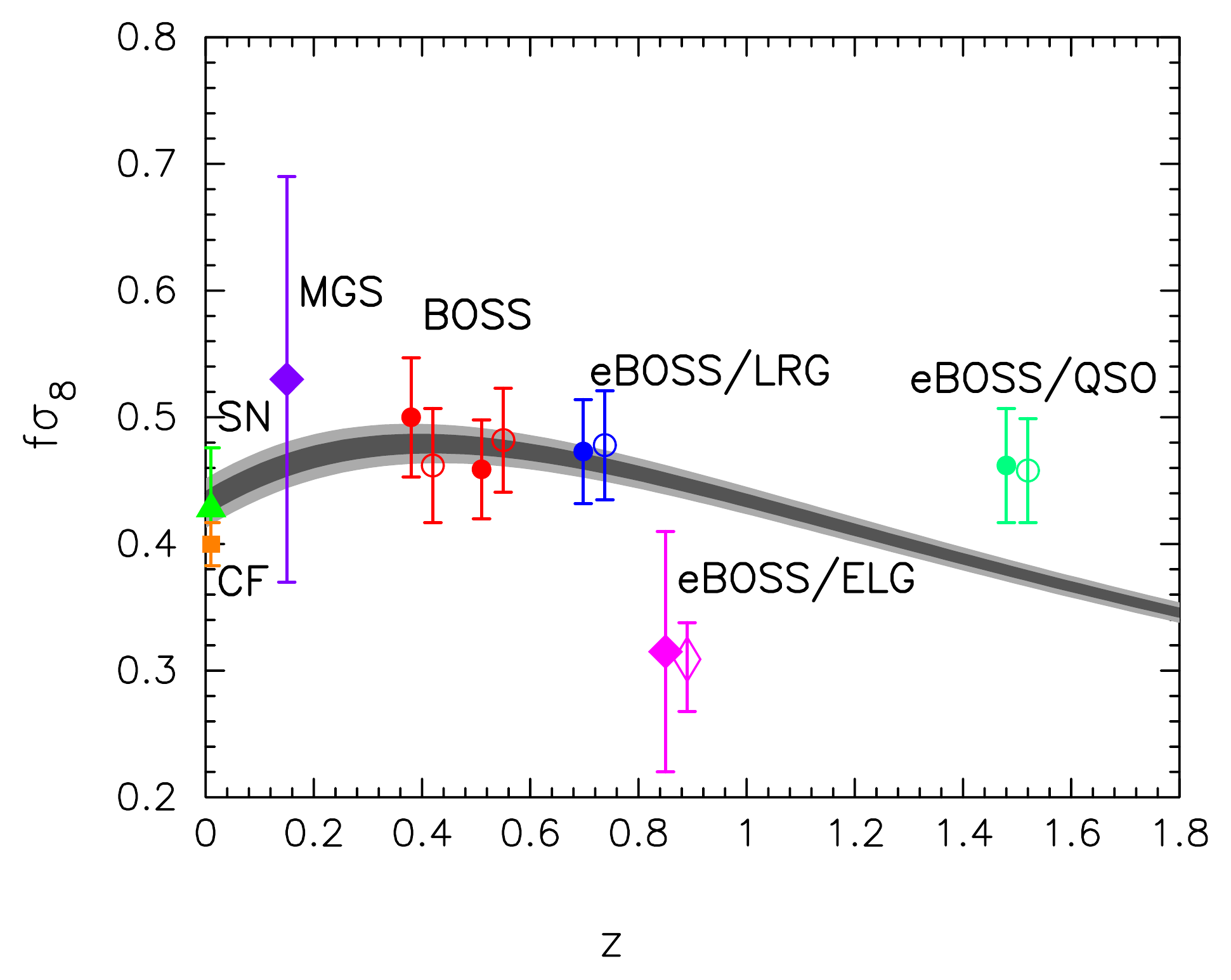} 
	\caption{Measurements of $f\sigma_8$ from various surveys. The green triangle at $z_{\rm eff} = 0.02$ is from an analysis of low redshift Type Ia SN by \citet{Huterer:2017} and the orange square is from
	the cosmic flows analysis of \citet{Boruah:2020} using SN, surface brightness fluctuations and 
	Tully-Fisher distances to nearby galaxies.  The filled symbols
	are from the consensus results from the completed BOSS/eBOSS surveys \citep{Alam:2021}. The measurement at	$z=0.15$ is for the SDSS Main Galaxy Sample \citep{Howlett:2015}. The open circles show a reanalysis of the BOSS, eBOSS/LRG and eBOSS/QSO samples by \citet{Brieden:2022} using their \textit{ShapeFit} methodology. The open diamond shows results from an effective field theory analysis of the eBOSS Emission Line Galaxy sample by \citet{Ivanov:2021}. The grey bands show the $1\sigma$ and $2\sigma$ regions allowed by the \Planck base \LCDM cosmology.}
	\label{fig:fsigma8vz}
\end{figure}

Redshift space distortions (RSD) provide an independent way to measure the growth rate of fluctuations \citep{Kaiser:1987}. RSD provide a measurement of the parameter combination $f\sigma_8$, where $f$ is the logarithmic derivative of the linear growth rate $D$ with respect to the scale factor 
\begin{equation}
f = {{ \rm d \ ln} D \over { \rm d \ ln} a}, 
\end{equation}
and $a = (1+z)^{-1}$. In the \LCDM model, $f \approx
\Omega_{\rm m}(z)^{0.55}$ \citep{Lahav:1991}. RSD measurements from large galaxy redshift surveys are  of particular relevance to  the $S_8$ tension because they
overlap with the redshift range covered by the KiDS survey.   Since most of the information on RSD from the analyses discussed in this section 
comes from wavenumbers in the range $0.01 h{\rm Mpc}^{-1}  \simlt k \simlt 0.1 h{\rm Mpc}^{-1}$
(see, for example, Fig. 8 of \citealt{Philcox:2022}) corresponding to linear or very mildly non-linear  scales, any discrepancy with the \Planck\ \LCDM\ model would be fatal to the explanation of the $S_8$ tension proposed in this paper. It is therefore important to review the consistency between RSD measurements and the  \textit{Planck} \LCDM model in detail.

As noted in \cite{Params:2016},  early RSD measurements used a fixed fiducial cosmology that differed significantly from the \textit{Planck} \LCDM cosmology  leading to potential biases and underestimates of the error on $f\sigma_8$ \citep{Howlett:2015}. In addition, there have been significant  theoretical advances in the modelling of RSD in recent years. To compute the RSD constraints in Fig.~\ref{fig:S8} we have used the results  shown by the 
filled symbols in Fig. \ref{fig:fsigma8vz}. The measurements
at $z_{\rm eff} = 0.38$ and $0.61$  are  the `consensus' results
from the completed  Baryon Oscillation
Spectroscopy Survey (BOSS) and extended BOSS (eBOSS) surveys \citep{Alam:2017, Alam:2021}\footnote{SDSS: \url{https://www.sdss.org/science/final-bao-and-rsd/-measurements-table}} supplemented by the 
the measurement from the  SDSS Main Galaxy Sample (MGS) at $z_{\rm eff} = 0.15$ \citep{Ross:2015, Howlett:2015} and the low
redshift ($z_{\rm eff} \approx  0.02$)  measurements of $f\sigma_8$ from \cite{Huterer:2017} and  \cite{Boruah:2020} from peculiar velocity-density field correlations (which we loosely bracket under the term 'RSD' in this section). The grey contours in Fig. \ref{fig:fsigma8vz} show the $1\sigma$ and $2\sigma$ ranges contours for the \textit{Planck} base \LCDM cosmology.

The pink contours in Fig. \ref{fig:S8} show the constraints in the $S_8$-$\Omega_{\rm m}$ plane derived by combining these RSD measurements (including the BAO constraints for the BOSS, eBOSS and MGS samples)
with the Pantheon supernova magnitude-redshift relation \citep{Scolnic:2017},
 Ly$\alpha$-quasar and Ly$\alpha$-Ly$\alpha$ 
BAO measurements from \citep{Blomqvist:2019,deSainteAgathe:2019} and  BAO measurements from the 6dF Galaxy Survey \cite{Beutler:2011}. To scale the BAO constraints we impose a Gaussian \textit{Planck} prior on the sound horizon,
 $r_{\rm d} = 147.31 \pm 0.31 \ \Mpc$.  
The RSD constraints in Fig. \ref{fig:S8} are consistent with the 
$S_8$ results from  \Planck TTTEEE. There is also
substantial overlap between the RSD and KiDS contours in Fig. \ref{fig:S8}.

The consensus RSD measurements plotted in Fig.~\ref{fig:fsigma8vz} assume a \Planck \LCDM fiducial power spectrum with adjustable BAO location parameters in addition to the velocity fluctuation amplitude $f\sigma_8$. Recently, several groups have developed `full-shape' analyses based on effective field theory (EFT) descriptions of non-linear perturbations \citep[e.g.][]{dAmico:2020, Ivanov:2020, Chen:2022, Philcox:2022, DAmico:2022}.
The EFT analyses aim to constrain cosmological parameters independently of \textit{Planck}. However, the nuisance parameters required to model perturbation theory counter-terms, galaxy biasing, and redshift space distortions,  effectively down-weight information at wavenumbers $k \simgt 0.2 h^{-1} {\rm Mpc}$. As a consequence of the restricted wavenumber range,  the primordial spectral index $n_{\rm s}$ is poorly constrained in comparison to \textit{Planck}. Constraining $n_{\rm s}$ to the \textit{Planck} best fit value,
\cite{Ivanov:2020} find $\sigma_8 = 0.721 \pm 0.43$, \cite{Philcox:2022} find $\sigma_8 = 0.729 \pm 0.040$ and \cite{Chen:2022} find $\sigma_8 = 0.738 \pm 0.048$, from fits to the BOSS power spectra. These are consistent with each other, but sit low compared to the \Planck \LCDM estimate,  $\sigma_8 = 0.8095 \pm 0.0074$, by $1.6 - 2 \sigma$ (see also \cite{Troster:2020}). The recent EFT analysis  of the BOSS power spectrum and bispectrum 
\citep{DAmico:2022} finds $\sigma_8 = 0.794 \pm 0.037$ (labelled as $P_\ell + B_0^{1{\rm loop}} + B_2^{\rm tree}$ in their analysis) in good agreement with both the \textit{Planck} \LCDM value, as well as the cosmic shear results\footnote{Note that \citet{DAmico:2022} apply a correction for  prior volume effects in the EFT analysis that bias $\sigma_8$ low by about $1\sigma$ if left uncorrected.}. Constraints in the $S_8-\Omega_{\rm m}$ plane from this analysis are shown by the peach contours in Fig. \ref{fig:S8}. 

Recently, \cite{Brieden:2021a,Brieden:2021b} have developed an extension of the `classical' BOSS/eBOSS analyses, by including an additional parameter that is sensitive to the parameter combination $\Omega_{\rm m} h$. The constraints on $f\sigma_8$ derived by applying their {\it ShapeFit} technique to the BOSS, eBOSS/LRG and eBOSS/QSO samples
\citep{Brieden:2022} are shown by the open symbols in  Fig. \ref{fig:fsigma8vz}. These agree almost perfectly with the consensus results from the BOSS/eBOSS collaborations reported in \cite{Alam:2017, Alam:2021}.  

The open diamond in Fig. \ref{fig:fsigma8vz}  shows the constraint $f\sigma_8 (z_{\rm eff} = 0.85) = 0.309^{+0.029}_{-0.041}$
from a full shape EFT analysis of the eBOSS/ELG sample by \cite{Ivanov:2021}. This estimate is in strong disagreement  with the \Planck base \LCDM value (by $\sim 4.7\sigma$)  and also  disagrees with the \LCDM\ parameters determined from the BOSS DR12 sample (see their Fig. 1). The \cite{Ivanov:2021} eBOSS/ELG estimate of $f\sigma_8$ is clearly an outlier in Fig.~\ref{fig:fsigma8vz}, suggesting that further work is required to 
establish whether the shape of the power spectra  is robust
to the large corrections required to account 
for selection biases in the eBOSS/ELG sample.

Evidently, there are still small  methodological differences in the RSD analyses that  lead to differences of $\sim 1-2\sigma$  in measurements of $f\sigma_8$ inferred from the same data. Although the pink contours in Fig. \ref{fig:S8} are consistent with the constraints from \Planck, they 
have substantial overlap with the KiDS and DES contours. RSD measurements
cannot yet distinguish decisively between the \textit{Planck} and weak lensing 
amplitudes \citep{Efstathiou:2018, Nunes:2021}. There is therefore no compelling evidence from RSD to support claims that linear growth rates are suppressed at redshifts $z \simlt 1$ compared to the expectations of  \LCDM \citep{Macaulay:2013, Nesseris:2017, Kazantzidis:2018, Benisty:2021, Snowmass:2022}.

Note also that the functional form of the linear growth rate $D(t)$ over the redshift range $\sim 0.2 - 0.7$ inferred from a tomographic analysis of cosmic shear, galaxy clustering and CMB lensing appears to be compatible with that expected in the  \LCDM model \citep{Garcia-Garcia:2021}, though the overall fluctuation amplitude  is found to be
lower than expected in the \textit{Planck} base \LCDM cosmology.

\subsection{CMB lensing}

Rather than reflecting anomalies in the  growth rate of
perturbations, the $S_8$ tension may be related to  
anomalies in the propagation of photons,  for example a departure from General Relativity that
results in a `gravitational slip' \citep[e.g.][]{Daniel:2008, Bertschinger:2011, Simpson:2013, Pizzuti:2019}.
This can be tested using CMB gravitational lensing. The CMB lensing signal is caused by matter distributed along the line-of-sight over a broad redshift range with a median redshift of $z \sim 2$.
This is not very much higher than the mean redshift $z\sim 1$ of the highest tomographic redshift bin in KiDS-1000 (which 
contributes  the highest weight to the KiDS two point statistics).

As noted in \cite{Plensing:2020}  for base \LCDM  with `lenspriors'
imposed\footnote{These are loose priors of $n_s = 0.96 \pm 0.02$, $\Omega_b h^2 = 0.0222 \pm 0.005$
(motivated by the deuterium abundance \citep{Cooke:2018}), $0.4 < h < 1$, and the optical depth to reionization fixed at 
$\tau= 0.055$.}, the \textit{Planck} lensing likelihood tightly constrains the
parameter combination\footnote{As this paper was nearing completion \cite{Carron:2022} presented results from an improved \Planck\ lensing likelihood.
This gives $\sigma_8 \Omega_{\rm m}^{0.25} = 0.599 \pm 0.016$, within $0.4\sigma$ of the \Planck\ TTTEEE result of Eq. \ref{equ:lens2}.}
\begin{subequations}
\begin{equation}
\sigma_8 \Omega_{\rm m}^{0.25} = 0.589 \pm 0.020,  \quad {\rm Plens},   \label{equ:lens1}
\end{equation}
which is within $1\sigma$ of the constraint derived from  the TTTEEE likelihood,
\begin{equation}
\sigma_8 \Omega_{\rm m}^{0.25} = 0.6057 \pm 0.0081.   \quad {\rm  TTEEE}. \label{equ:lens2}
\end{equation}
\end{subequations}

The addition of BAO data and a conservative prior of $\theta_{\rm MC} =
1.0409 \pm 0.0006$\footnote{Where $\theta_{\rm MC}$ is an
  approximation to the angular size of the sound horizon at
  recombination, see \cite{Params:2016} for a definition.}
(effectively adding another BAO point at recombination) to the
\textit{Planck} lensing likelihood restricts the range of allowed values of
$\Omega_{\rm m}$ resulting in  the green 
contours\footnote{Computed from the chains
  `base$\_$lenspriors$\_$BAO$\_$theta' available from the
  \textit{Planck} Legacy Archive ({\tt https://www.cosmos.esa.int/web/planck/pla}).} in Fig.~\ref{fig:S8}. Assuming base
\LCDMns, the \textit{Planck} lensing constraints are in very good agreement
with the \textit{Planck} TTTEEE results (as they should be, since acoustic
peaks of the temperature and polarization power spectra are sensitive
to CMB lensing, \citealt[see e.g.][]{Larsen:2016}). Over the redshift range
probed by \textit{Planck} lensing, there is no evidence from Fig.~\ref{fig:S8} for any gravitational slip/modified gravity type of effect that might be causing photons to behave in a different way from the predictions of \LCDMns.

It is possible to test consistency of weak galaxy and CMB lensing at redshifts $z \simlt 1$  by cross-correlating CMB lensing  with weak lensing maps and/or maps of the galaxy distribution \citep[e.g.][]{Marques:2020, Robertson:2021, Krolewski:2021, Chang:2022, ChenWhite:2022}.
We will discuss these measurements briefly in Sect. \ref{sec:conclusions}.

\begin{figure*}
	\centering
	\includegraphics[width=\textwidth, angle=0]{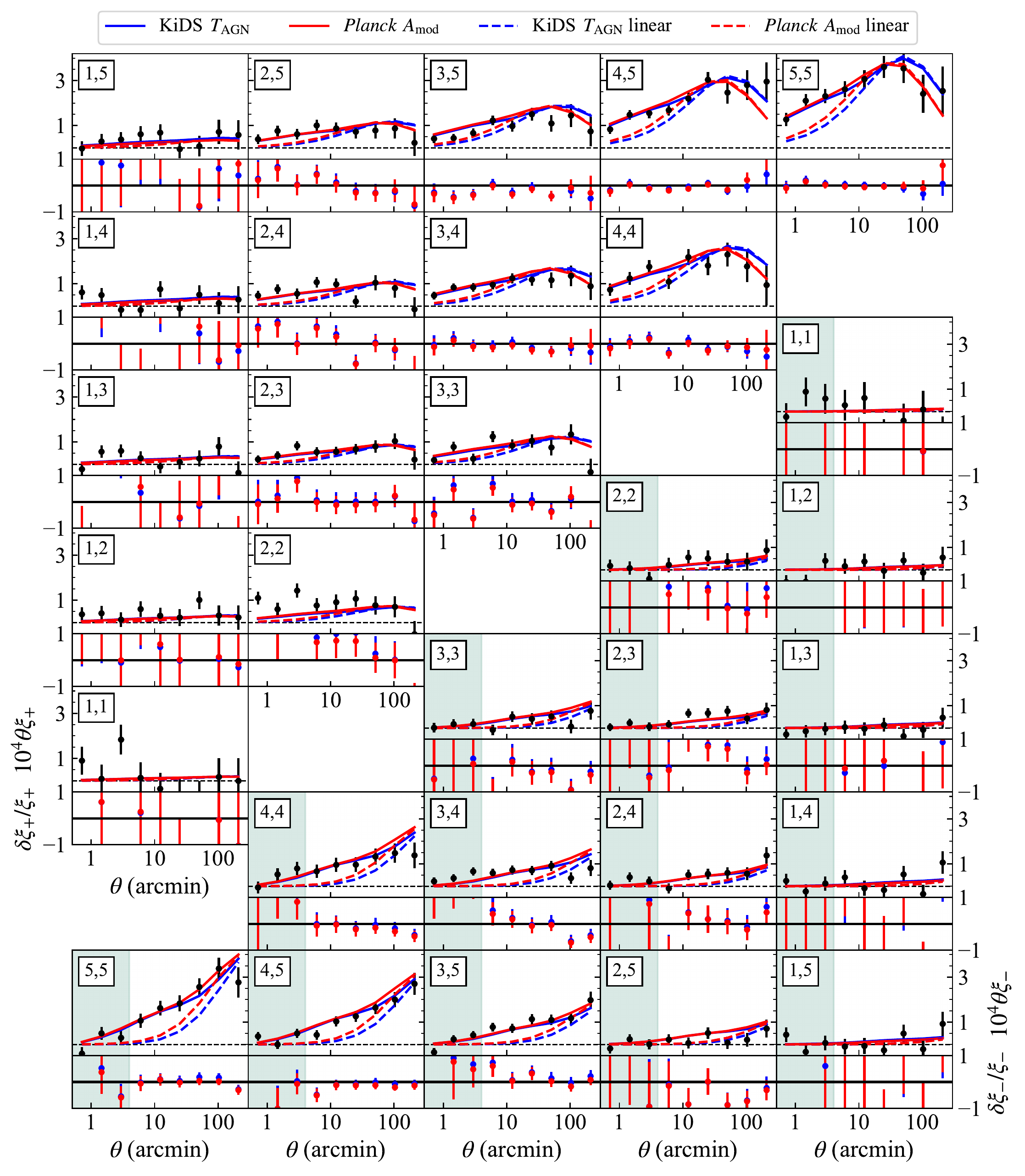} 
	\caption{Measurements of KiDS21 two-point correlation functions $\xi_{\pm}$, scaled by the angular separation, $\theta$ (upper panels).  The correlation functions are measured for each redshift bin pair, indicated by the label and the error bar represents the square root of the diagonal of the analytic covariance matrix. The blue line is the best fit \LCDM theoretical prediction, including modelling of the non-linear matter power spectrum using {\textsc HMCode2020} (very similar to the KiDS21 fit, which used {\textsc HMCode2016}), which includes a baryon feedback parameter $\Theta_{\rm AGN}$. The red line fixes cosmological parameters to the \Planck best-fit values and now modulates the non-linear power spectrum using the parameter $A_{\rm mod}$ (Eq.~\ref{equ:NL}).  In both cases, the dashed lines show $\xi_{\pm}$ computed from the linear matter power spectra. The lower panels show the residual differences between the models and the KiDS21 measurements $\delta\xi_{\pm}/\xi_{\pm}$. Following KiDS21, scales excluded from the analysis are shaded in green.}
	\label{fig:xipm_kids}
\end{figure*}

\section{A possible solution}
\label{sec:solution}

\subsection{Motivation}
\label{subsec:motivation}

\begin{table*}
  \centering
  \caption{Mean value of $S_8$ and 1$\sigma$ errors from our
    The entry
   labelled {\sc HMCode2016} uses the identical  non-linear matter power spectrum  and feedback model as in KiDS21
   and reproduces their results to high accuracy. The next set of  entries use {\sc HMCode2020} to  model the non-linear power spectrum
   applying uniform priors with  different ranges for the baryon feedback parameter $\Theta_{\rm AGN}$. (The row labelled `no feedback' has all baryon feedback switched off.) The asterisk denotes the case of a narrow prior on $\Theta_{\rm AGN}$, as assumed by \citet{Troester:2021extended}, which we refer to as the `fiducial' model in this paper. The final set of entries show results for the phenomenological non-linear suppression model governed by the parameter $A_{\rm mod}$ (see Eq.\ref{equ:NL}).   For the cases labelled `free', the  cosmological parameters are allowed to vary freely. For cases labelled `\Planck' the cosmological parameters are fixed to the TTTEEE values for the \Planck best fit \LCDM cosmology. The column labelled $N_\sigma$ lists  $(\chi^2_{\rm min}- \sqrt{2 N_{\rm deg}})/\sqrt{2 N_{\rm deg}}$  where $N_{\rm deg}$ is the effective number of degrees of freedom. We have $N_{\rm deg} =
   225 - 4.5$ (accounting for the number of constrained  parameters in the fits following \citealt{Joachimi:2021}) for the `free' fits where cosmological parameters are allowed to vary in addition to nuisance parameters (see KiDS21) and $N_{\rm deg} =
   225 - 2.5$ for the fits with cosmological parameters constrained to the \textit{Planck} values. 
   The column labelled PTE gives the probability to exceed the value of $\chi^2_{\rm min}$.
   }
\begin{tabular}{@{}lccccc}
\hline
Non-linear model \& prior range  & Cosmology &  $S_8$ & $\chi^2_{\rm min}$ & $N_\sigma$ & PTE \\
\hline
     
       \ {\sc HMCode2016}  & free &  $0.765\pm{0.018}$ & 260.1 &$1.92$& $0.030$ \\ 
      
\hline 
      \ {\sc HMCode2020} no feedback & free & $0.755\pm{0.016}$ &  261.5 & $1.95$ & $0.025$ \\  
       *{\sc HMCode2020} $\Theta_{\rm AGN}=7.3-8.3$  & free & $0.774 \pm{0.021}$ & 260.2 & $1.89$ & $0.029$ \\ 
       \ {\sc HMCode2020} $\Theta_{\rm AGN}=7.0-10.0$ & free & $0.785\pm{0.030}$ & 260.0 &  $1.88$ & $0.030$\\
       \ {\sc HMCode2020} $\Theta_{\rm AGN}=7.0-10.0$ & \Planck & $0.829$ & 267.6 & $2.13$ & $0.016$ \\
      
       \hline
       \ {\sc HMCode2020} $A_{\rm mod}=0.5-1.2$ & free & $0.780\pm{0.035}$ & 260.3 & $1.89$ & $0.029$ \\
       \ {\sc HMCode2020} $A_{\rm mod}=0.5-1.2$ & \Planck  & $0.829$ & 265.5 & $2.04$ & $0.021$ \\ 
      
\hline
\end{tabular}
\label{tab:chi2}
\end{table*}

To summarize the previous section:

\smallskip

\noindent
(i) There is evidence from independent  weak lensing surveys  that the amplitude of the matter fluctuations, as measured by the $S_8$ parameter, is lower than expected according to the \textit{Planck} \LCDM cosmology at about the $2.5-3\sigma$ level.

\smallskip

\noindent
(ii) Redshift space distortions analyses over the redshift range $0.02 - 1.6$ are consistent with the \textit{Planck} base \LCDM cosmology, suggesting that perturbations grow at the rates expected in \LCDM over this
entire redshift range. 

\smallskip

\noindent
(iii) CMB lensing is consistent with the expectations of the 
 \textit{Planck} base \LCDM cosmology, suggesting that photons behave as expected in this cosmology and that perturbations grow at the expected rates over the redshift range $z\sim 1000$  to $z\simlt 2$.

\smallskip

\noindent
(iv)  In addition, the combination of BAO measurements and magnitude-redshift relation of Type 1a supernovae  tightly constrains the expansion history $H(z)$ to
be very close to that of the \textit{Planck} base \LCDM cosmology  over the redshift range probed by the weak lensing surveys \citep{Heavens:2014, Verde:2017, Macaulay:2019, Efstathiou:2021}. Solutions to the $S_8$ tension involving a
change in the background cosmology (e.g. invoking a late time transition to a phantom equation of state as considered by \citealt{Joudaki:2017}) are disfavoured by the BAO and supernova data.

\smallskip

In this paper we will  seek a physical explanation of the $S_8$ tension that ties together points (i)-(iv).

Figure~\ref{fig:xipm_kids} shows the KiDS $\xi_{\rm \pm}$ measurements as reported in KiDS21. The shaded regions in the $\xi_{-}$ plot shows the range of  angular scales  that are excluded from the cosmological analysis by KiDS21 to reduce uncertainties associated with the modelling of baryonic feedback. We use the public KiDS pipeline and follow the updated analysis of \citet{Troester:2021extended}, who re-analyse the KiDS21 band-power measurements with {\sc HMCode2020} \citep{Mead:2021} in place of the earlier version of {\sc HMCode} \citep{Mead:2015} used in KiDS21. {\sc HMCode2020} provides a  description of the non-linear evolution of the matter power spectrum and models baryonic feedback via a parameter
$\Theta_{\rm AGN} = {\rm log}_{10}(T_{\rm AGN}/{\rm K})$ calibrated using the {\sc BAHAMAS} (BAryons and HAloes of MAssive Systems) hydrodynamical simulations \citep{McCarthy:2017,vandaalen:2020},  where $T_{\rm AGN}$ is a subgrid heating parameter. The solid blue line in this figure shows our best fit to these data applying the same priors on parameters, including $\Theta_{\rm AGN}$, as in \citep[][see their Appendix A]{Troester:2021extended}. The marginalised mean value of the $S_8$ posterior distribution\footnote{Throughout this paper we always report the marginalised mean value of the posterior distribution and the 68\% confidence limits.} is quoted in Table~\ref{tab:chi2} (entry labelled  {\sc HMCode2020} $\Theta_{\rm AGN}=7.3-8.3$, which we will refer to as the `fiducial' analysis). This is within $0.5\sigma$ of the analysis reported in Table~\ref{tab:chi2} using {\sc HMCode2016} (which agrees perfectly with the results of KiDS21), consistent with the impact on the band powers statistics as reported by \citet[][Appendix A]{Troester:2021extended}.
The cosmological parameters for this fit will be discussed further in Sect.~\ref{subsec:TAGN}.
 
  The  blue dashed lines in Fig.~\ref{fig:xipm_kids}  show the predictions for the same parameters but now using only the {\it linear}  theory matter power spectrum. One can see that the non-linear corrections make a dominant contribution to the total signal, even for $\xi_{+}$ except on angular scales $\simgt 10-20$ arcmin. Evidently, extracting precision cosmology from smaller angular scales requires accurate modelling of the non-linear power spectrum.
 
Given this sensitivity, we investigate the plausibility of solving the $S_8$ tension 
by modifying the spectrum only in the non-linear regime, while fixing the linear power spectrum to that of the \textit{Planck} base \LCDM cosmology.  
With such a solution it may possible to explain the cosmic shear
measurements in a way that is consistent with points (i)-(iv) above, since the RSD measurements and \textit{Planck} CMB lensing are insensitive to highly non-linear scales.  In addition, such a solution would not require any modification to the \textit{Planck} \LCDM expansion history, preserving the consistency of the \LCDM\ model with the magnitude-redshift relation of Type Ia supernovae, and the geometrical constraints from BAO measurements.

\subsection{Phenomenological Model}
\label{subsec:Amod}

The lensing data are not yet  sufficiently precise to allow a parametric reconstruction of the matter power spectrum as a function of redshift, so in this section we adopt a particularly simple phenomenological  model. At this stage,
we wish to investigate whether a model with the \textit{Planck} \LCDM cosmological parameters and linear fluctuation spectrum can provide an acceptable fit to the KiDS $\xi_{\pm}$ measurements via a modification of the spectrum in the non-linear regime. We write the matter power spectrum, $P_{\rm m}(k, z)$, as
\begin{equation}
P_{\rm m}(k, z) =  P^{\rm L}_{\rm m}(k, z) + A_{\rm mod}[P^{\rm NL}_{\rm m} (k, z) - P^{\rm L}_{\rm m}(k, z)] \,, \label{equ:NL}
\end{equation} 
where the superscript  ${\rm L}$ denotes the linear theory power spectrum, NL denotes the non-linear power spectrum computed by {\sc HMCode2020} with baryonic feedback switched off and $A_{\rm mod}$ is an amplitude parameter that we vary in the MCMC analysis.   

\begin{figure}
	\centering
	\includegraphics[width=\columnwidth]{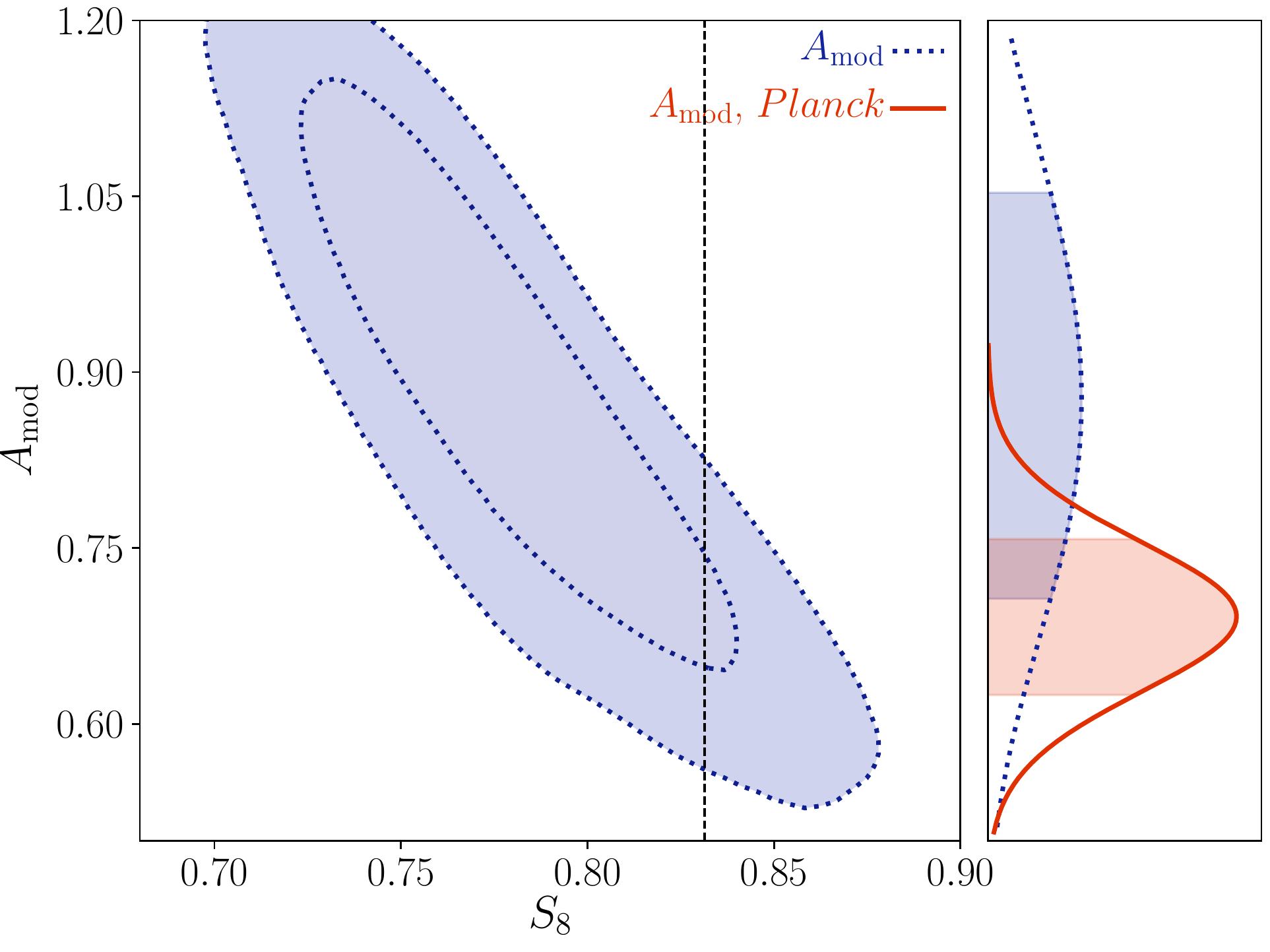} 
	\caption{Illustration of the strong degeneracy between $S_8$ and
	the phenomenological power spectrum suppression parameter $A_{\rm mod}$. The blue contours shown in the left-hand panel represent the 
	68\% and 95\% constraints for the KIDS $\xi_{\pm}$ statistics analyzed by using {\sc HMCode2020}-no feedback, but instead varying $A_{\rm mod}$. The dashed line indicates the \Planck \LCDM best-fit value for the $S_8$ parameter.  The posterior of $A_{\rm mod}$ for this case is shown by the blue 
	curve in the right hand panel. If the cosmological parameters are fixed to the  \Planck \LCDM best-fit values we find the posterior for 
	$A_{\rm mod}$  shown by the red curve in the right hand panel.
	}
	\label{fig:Amod}
\end{figure}

Fixing the cosmology to the \Planck TTTEEE \LCDM parameters, but allowing the `nuisance' parameters describing  intrinsic alignments, redshift calibration errors and 
$A_{\rm mod}$ to vary, we find the best fit theory model shown by the solid red  line in Fig.~\ref{fig:xipm_kids}. As can be seen, this model is almost indistinguishable from the blue line of the KiDS cosmology. The
posterior distribution of the parameter $A_{\rm mod}$ is shown in red in Fig.~\ref{fig:Amod}. We find a mean
\begin{equation}
      A_{\rm mod} = 0.69\pm0.06 \label{equ:amp} \, ,
\end{equation}
and a minimum value of $\chi^2_{\rm min} = 265.5$ for $225$ data points,  which is very close
to the minimum value of $\chi^2$ for the fiducial model (see Table \ref{tab:chi2}). Both models provide 
acceptable fits to the data. (All of the $\chi^2$ values listed in Table \ref{tab:chi2} are high by
about $2\sigma$ as a consequence of outliers, for example, the points at $\theta \simlt 50^\prime$ in the
(2,2) $\xi_+$ correlation function).   Note that the uncertainty in Eq.~\ref{equ:amp} is  an underestimate as we have neglected the error on the \Planck base \LCDM cosmology. The reanalysis of the KiDS-1000 $\xi_{\pm}$ measurements varying both the $A_{\rm mod}$ and cosmological parameters is shown as the blue contour in Fig.~\ref{fig:Amod}, highlighting the degeneracy between $S_8$ and $A_{\rm mod}$, which reduces the power of lensing data to constrain $S_8$. 

We find that fixing the cosmology to Planck does not cause any significant shifts to the
nuisance parameters compared to the fiducial analysis. Similarly, we have verified that  fixing the nuisance parameters to the best fit values of the fiducial analysis constraints does not 
alter the constraint on $A_{\rm mod}$. These tests show that the low value of $A_{\rm mod}$ in Eq. \ref{equ:amp} is insensitive to the nuisance parameters.

The results of this section show that a suppression of the non-linear power spectrum can  reconcile the base \Planck cosmology with the KiDS lensing data. However, the model is purely phenomenological and so it is reasonable to ask whether such a suppression has a physical interpretation. There are two obvious possibilities: (a) that the suppression is caused by baryonic feedback; (b) that it is caused by a physical property of the dark matter that suppresses the power spectrum on small scales, for example some type of self-interaction \citep[see e.g. the review by][]{Tulin:2018} or other non-gravitational interaction \citep[e.g.][]{Becker:2021}, a mixture
of cold and warm dark matter \citep[e.g.][]{Boyarsky:2009}, or an axionic dark matter component with a de Broglie wavelength of a few Mpc \citep[e.g.][]{Hui:2017}. We consider baryonic feedback in the next subsection.

\subsection{Baryonic Feedback}\label{subsec:TAGN}

\begin{figure*}
	\centering
	\includegraphics[width=\textwidth]{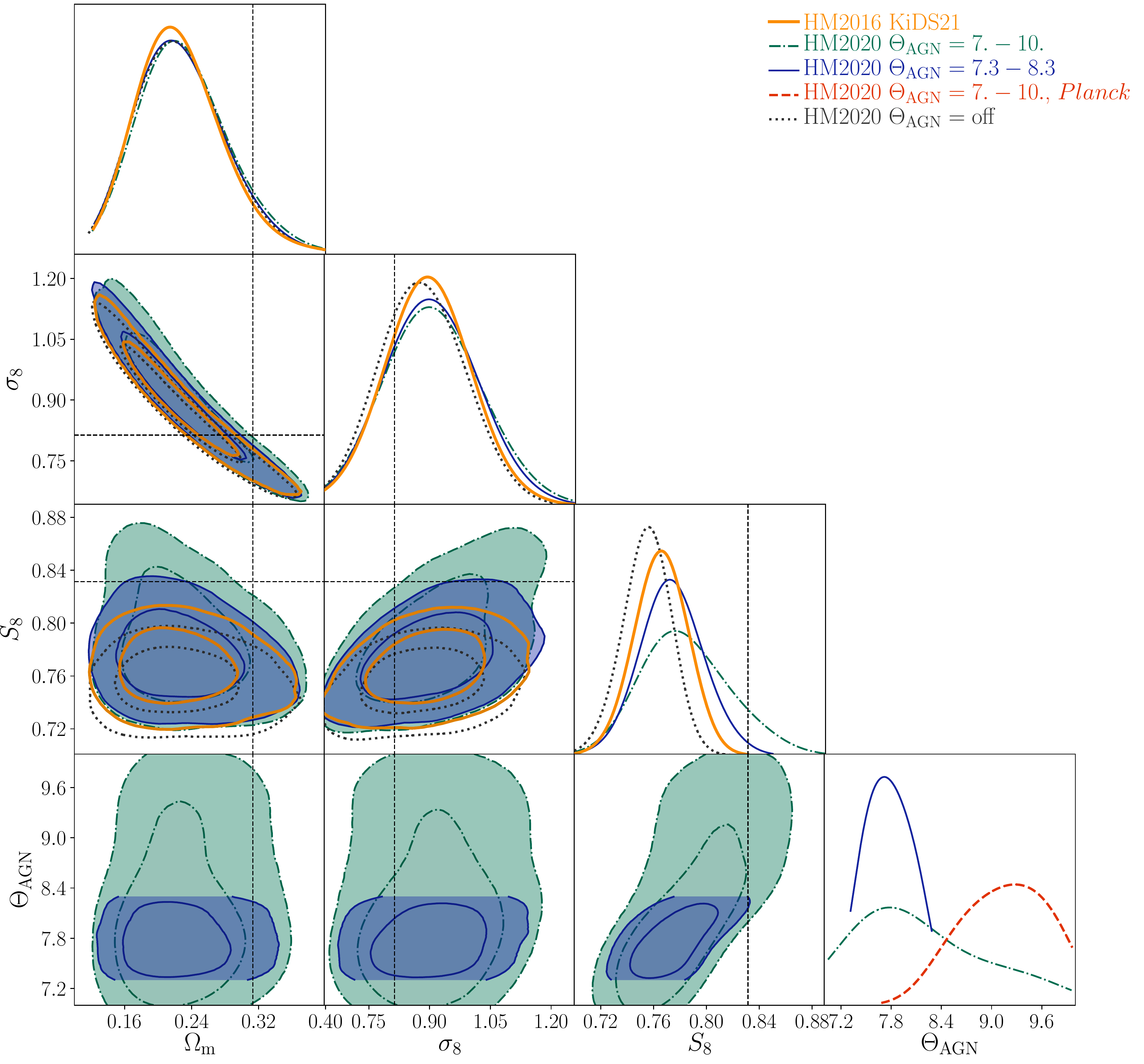} 
	\caption{68\% and 95\% constraints for KIDS $\xi_{\pm}$ statistics analyzed using {\sc HMCode2020}, with varying prior ranges for 
	$\Theta_{\rm AGN}$, compared to the HM2016 analysis by KiDS21 (yellow). Following \citet{Troester:2021extended}, we allow $\Theta_{\rm AGN}$ to vary with a uniform prior over the range $7.3-8.3$ (blue). The range is extended to $7.1-10.0$ (green) to allow for more extreme feedback. 
	The constraints neglecting feedback are shown in black dotted. We fix the cosmological parameters to \Planck \LCDM\ best fit values and find the posterior on $\Theta_{\rm AGN}$ shown by the dashed red line in the bottom right panel. The dashed lines in the
	other panels show  the \Planck \LCDM best-fit values for $\Omega_{\rm m}$, $\sigma_8$ and $S_8$. }
	\label{fig:TAGN}
\end{figure*}

Hydrodynamic simulations of structure formation have improved remarkably over the last decade \citep[e.g.][]{Schaye:2015,   McCarthy:2017, Nelson:2019}. Nevertheless, the detailed physics involved in baryon  cooling, star formation, stellar feedback  and feedback from 
active galactic nuclei (AGN) remains extremely uncertain.  
As pointed out by many authors, feedback process redistribute the baryons leading to a suppression of the dark matter power spectrum in the non-linear regime \citep[see e.g.][and references therein]{vandaalen:2011, Vogelsberger:2014, Hellwing:2016, Chisari:2019}. Since feedback processes are not well understood, there is considerable uncertainty in the overall amplitude of this suppression, its scale dependence and evolution with redshift. In this section we use the feedback scheme implemented in 
 {\sc HMCode2020} and ask how far do we have to vary the feedback parameter $\Theta_{\rm AGN}$ to match the  suppression of our phenomenological model of  Eq.~\ref{equ:amp}.

 We first pin the cosmological parameters to the \Planck TTTEEE \LCDM parameters then fit the KiDS
 $\xi_{\pm}$ data using  {\sc HMCode2020} with a broad  prior\footnote{This test is  similar to that of \citet{Yoon:2020}, who use {\sc HMCode2016} and KiDS-450.} on 
 $\Theta_{\rm AGN}$ that is uniform in the range $7.0 < \Theta_{\rm AGN} < 10.0$   . This prior range is substantially wider than the range $7.6 < \Theta_{\rm AGN} < 8.0$ over which 
{\sc HMCode2020} was calibrated against  the BAHAMAS simulations. As we move outside the calibrated regime, the link between the  parameter $\Theta_{\rm AGN}$ and  physical models of  feedback  becomes more tenuous. 
 
 The posterior distribution of the $\Theta_{\rm AGN}$ parameter is shown in red dashed in Fig.~\ref{fig:TAGN} and has a marginalised mean of
\begin{equation}
      \Theta_{\rm AGN} = 9.17^{+0.2}_{-0.8} \, .
      \label {equ:tagnpl}
\end{equation}
As in the previous section, we have verified that fixing the cosmology to \textit{Planck} \LCDM\ does not lead to significant differences in the nuisance parameters compared to the fiducial analysis and  that fixing the
nuisance parameters to their fiducial best-fit values does not
change the constraint of Eq.~\ref{equ:tagnpl}. This model gives
a minimum value of $\chi^2_{\rm min} = 267.6$ (see Table~\ref{tab:chi2}), almost identical to the minimum value of $\chi^2$ for the  $A_{\rm mod}$ model with cosmological parameters fixed to \textit{Planck}.  Note that the error in Eq.~\ref{equ:tagnpl} does not include the uncertainty in the \textit{Planck} base \LCDM cosmology.  As expected, 
{\sc HMCode2020} with a broad prior on $\Theta_{\rm AGN}$ behaves qualitatively like the phenomenological $A_{\rm mod}$ model.

Next, we fit the $\xi_{\pm}$ measurements allowing the cosmological parameters to vary while extending the prior range for $\Theta_{\rm AGN}$ in {\sc HMCode2020}.    We have also performed a fit with baryonic feedback switched off. The parameter constraints for these fits are shown in  Fig.~\ref{fig:TAGN}. Table~\ref{tab:chi2} and gives values for  $S_8$ 
and $\chi^2$ for these fits. 
Evidently, the parameters are degenerate and opening up the prior
on $\Theta_{\rm AGN}$ leads to long tails extending to high values
of $S_8$.  Fig.~\ref{fig:TAGN} illustrates clearly how
high values of $S_8$ are disfavoured by the assumed prior on the  baryonic feedback model adopted by \citet{troester:2021}. 

\section{Comparison with the effects of baryonic feedback in numerical simulations}
\label{sec:interpretation}

\begin{figure*}
	\centering
	\includegraphics[width=\textwidth]{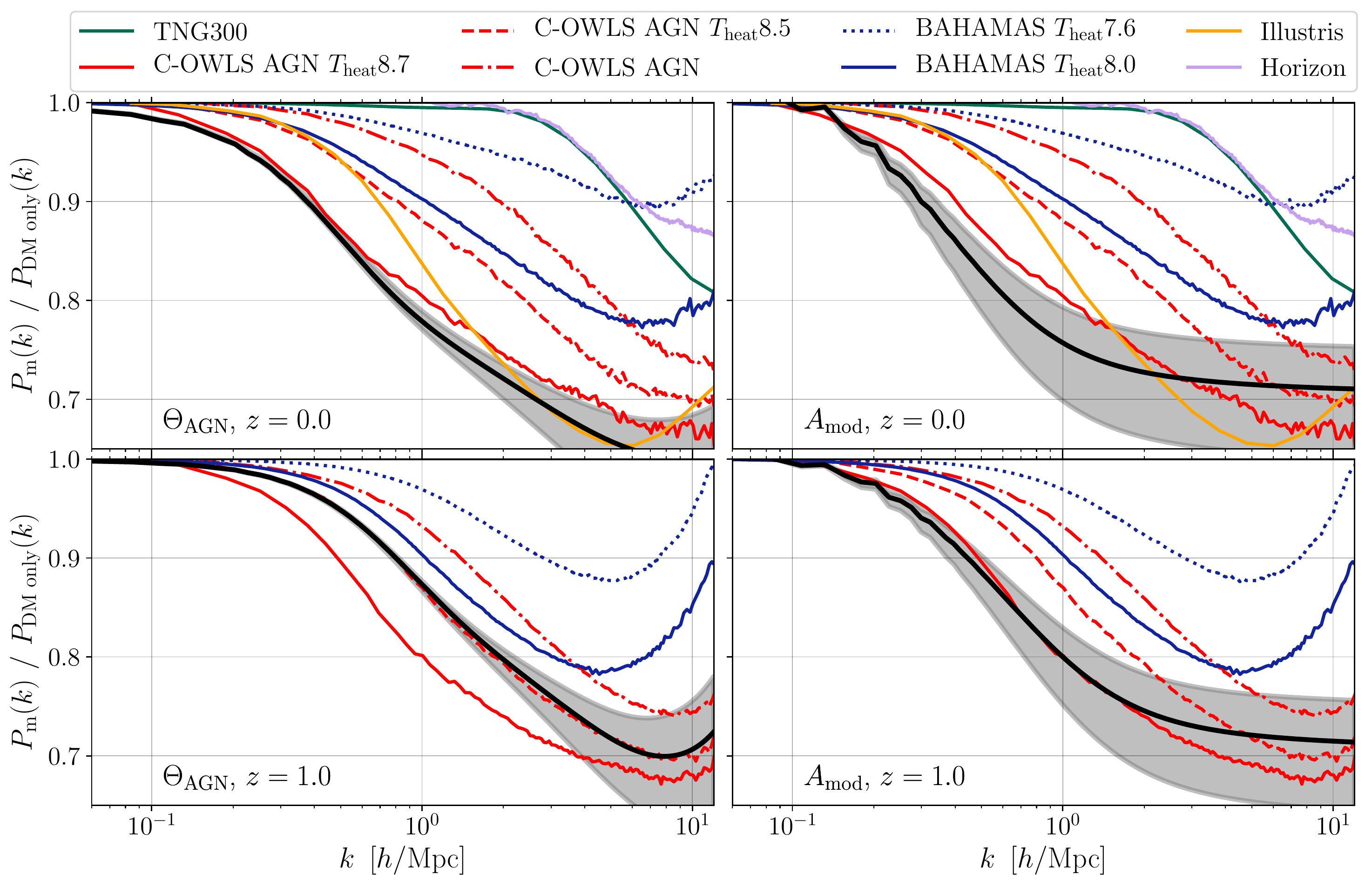} 
	\caption{\label{fig:baryons}
	The suppression of the matter power spectrum, $P_{\rm m}(k)/P_{\rm DM only}(k)$, required to match the KiDS $\xi_{\pm}$ measurements assuming the \Planck \LCDM cosmology. The upper and lower panels shows the suppression at $z=0$  and $z=1$ respectively. The left hand panels show the {\sc HMCode2020} model of Eq. \ref{equ:tagnpl} and the right hand panel shows the phenomenological model of Eq. \ref{equ:amp}.
	 The grey bands show the $1\sigma$ allowed ranges. The coloured lines, taken from \citet{vandaalen:2020}, show the suppression measured from cosmological hydrodynamical simulations incorporating baryonic feedback. The sources
	 are as follows: Illustris \citep[yellow; ][]{Vogelsberger:2014b}, Illustris TNG300 \citep[green; ][]{Springel:2018},  Horizon \citep[purple;][]{Dubois:2014}, C-OWLS \citep[red; COSMO-Overwhelmingly Large Simulations][]{LeBrun:2014} and {\sc BAHAMAS} \citep[blue;][]{McCarthy:2017} for several values of their AGN subgrid heating parameters, ${\rm log}_{10} (\Delta T_{\rm heat}/{\rm K})$.}

\end{figure*}

Very little is known  about the dark matter in the Universe. Therefore, it is  possible that a resolution of the $S_8$ tension along the
lines of our $A_{\rm mod}$ parameter is reflective of  new physics in the dark sector. However, as the landscape of such new physics is enormous and highly speculative, it is reasonable to investigate first whether the non-linear power suppression required to resolve the $S_8$ tension can be realized using less exotic physics. 
The posteriors of the $\Theta_{\rm AGN}$ parameter in Fig.~\ref{fig:TAGN} show that the KiDS weak lensing data alone are not able to constrain both  cosmological parameters and the {\sc HMCode2020}  baryonic feedback model. Additional information is therefore necessary to  quantify the effects of baryonic feedback on cosmic shear.
In this Section we discuss whether the non-linear power spectrum suppression required to reconcile the \textit{Planck} \LCDM\ cosmology with KiDS weak lensing is compatible with hydrodynamic numerical simulations that include feedback processes.

The  power spectrum suppression required by Eq.~\ref{equ:tagnpl} is shown 
by the heavy black lines on the the left-hand side of Fig.~\ref{fig:baryons}, together with $1\sigma$ error bands.  The suppression required by the phenomenological model, Eq. \ref{equ:amp}, is shown in the right-hand panels.  The various coloured lines show the suppression of the matter power spectrum caused by baryonic feedback as predicted by a number of cosmological hydrodynamical simulations at $z=0$ and $z= 1$. The simulation results are from the
library\footnote{\url{https://powerlib.strw.leidenuniv.nl}} compiled by \citet{vandaalen:2020}, comprised of the original sources cited in the caption of Fig.~\ref{fig:baryons}. For both  models, the power spectrum suppression required to reconcile the KiDS cosmic shear measurements with the \Planck cosmology is more extreme than predicted by most of the hydrodynamic simulations.\footnote{The  DES22 \LCDMns-Optimised analysis shown in Fig.~\ref{fig:S8} gives a slightly higher value of $S_8$ compared to the KiDS analysis but is lower than the \textit{Planck} \LCDM\  value (as in all previous cosmic shear analyses). DES22 makes different analysis choices compared to KiDS21 {\it e.g.} to reduce sensitivity to baryonic feedback \citep{Krause:2022}. The DES approach is to apply more conservative angular scale cuts designed to mitigate feedback as predicted by C-OWLS AGN (red dot-dashed line in Fig.~\ref{fig:baryons}). A detailed analysis is therefore required to compare and combine the KiDS21 and DES22 results. Nevertheless it is likely that a less extreme non-linear correction is required to explain the DES $S_8$ tension compared to KiDS.}  
At $z=1$, only the C-OWLS simulations with high values of the AGN feedback parameter can match the suppression required in our models, though these models fail to reproduce the local gas fractions in groups and clusters \citep{McCarthy:2017}.

Is the power spectrum suppression  required to alleviate the $S_8$ tension too extreme for baryonic feedback?   \cite{Chisari:2019} provide a comprehensive review of the modelling of baryonic feedback mechanisms  in cosmological simulations and their effects on the matter power spectrum.
The fiducial BAHAMAS model adopts $\Theta_{\rm AGN} = 7.8$, since this value reproduces the gas mass fractions in groups and clusters and the galaxy stellar mass function at $z\sim 0.1$ \citep{McCarthy:2017}.  The prior
of $7.3 < \Theta_{\rm AGN} < 8.3$ used by \cite{troester:2021} is broader than the range over which {\sc HMCode2020} has been calibrated against the BAHAMAS simulations and is intended to place conservative bounds on physically reasonable models of feedback. 

However, as
noted by \cite{Chisari:2019} baryonic feedback processes are complex and poorly understood.
It is perhaps possible that  hydrodynamical simulations do not capture the complexities of 
feedback. For example, it is not feasible to model processes such as  
black hole growth, magnetic fields, jet formation, cosmic ray injection ab initio.
\citep[see e.g.][]{Ensslin:2011, Hopkins:2021,Beckmann:2022}.  It is also not clear whether
 simulations  calibrated on  the local gas fractions of groups and clusters can be extrapolated to 
 redshifts $ z \sim 0.5-1$ relevant to the interpretation of cosmic shear surveys.  It is therefore important to develop  empirical ways of constraining baryon feedback rather than relying on numerical simulations. 

Joint analyses and cross-correlations between cosmic shear and the kinetic and thermal Sunyaev-Zeldovich (SZ) effects \cite[e.g.][] {schneider:2021, troester:2021} offer a promising way of testing baryonic feedback.  In particular, SZ observations are sensitive to baryons on larger spatial scales than X-ray observations. 
\cite{schneider:2021} use the shear power spectrum measurements from KiDS21, Atacama Cosmology Telescope (ACT) measurements of the kinetic SZ \citep{Schaan:2021}, together with gas and stellar fractions in groups and clusters inferred from X-ray measurements  \citep{Giri:2021}
to reconstruct the non-linear suppression of power using their baryonification model of feedback. If they constrain the baryon fraction to be within the range allowed by the CMB, they find a power suppression (their Fig. 7) that is extreme than most of the simulations, including BAHAMAS. In particular, they find evidence for a significant  suppression of power at the wavenumbers $\simlt 1 \,h {\rm Mpc}^{-1}$ in good agreement with our best fit $A_{\rm mod}$ model. The cross-correlation analysis of \citet{troester:2021} of the \Planck thermal SZ and shear power spectrum measurements  supports high values of $\Theta_{\rm AGN}$ that are more extreme than predictions from {\sc BAHAMAS}, but are restricted by the prior range adopted for  $\Theta_{\rm AGN}$ (a similar effect can be seen in Fig.~\ref{fig:TAGN}).
Hints for  more extreme feedback have also been found from other analyses that utilise ACT SZ information \citet{Amodeo:2021, Pandey:2021}. 
 
To summarize, our analysis shows that a suppression of power extending to large scales corresponding to wavenumbers of $ k \sim 0.2 \ h {\rm Mpc}^{-1}$   is required if the \textit{Planck} \LCDM cosmology is to be reconciled with cosmic shear.  Such strong feedback on large scales is not seen in most of the  hydrodynamic simulations shown in Fig.~\ref{fig:baryons}.  If such extreme baryonic feedback can be ruled out, then a non-linear solution to the $S_8$ tension would probably require new dark matter physics.

\section{Conclusions}
\label{sec:conclusions}

This paper is motivated by the substantial evidence from galaxy lensing measurements that the parameter $S_8$ has a lower value than expected in the \textit{Planck} \LCDM cosmology.
Noting that the signal-to-noise driving the cosmic shear constraints is dominated by non-linear scales, we have investigated whether the \textit{Planck} \LCDM cosmology can be reconciled with these measurements by modifying the matter power spectrum on non-linear scales, preserving all other features of \LCDMns. If this explanation is correct:

\smallskip

\noindent
$\bullet$ growth rate measurements that are sensitive mainly to linear scales,  in particular, precision measurements of RSD from  wavenumbers $k \simlt 0.1 \,h {\rm Mpc}^{-1}$ using forthcoming data from the Dark Energy Spectroscopic Instrument \citep[DESI;][]{DESI:2016}, should agree with the \textit{Planck} cosmology.

\smallskip

\noindent
$\bullet$ the background expansion history, $H(z)$,  and consequently  
luminosity and angular diameter distances, should agree with the \textit{Planck} \LCDM\ cosmology at all redshifts; 

\smallskip

\noindent
$\bullet$ photons should respond to gravity as expected in General Relativity. Gravitational lensing measurements (based on  either the CMB or galaxies) that are sensitive primarily to linear scales,  should agree with the \textit{Planck} \LCDM cosmology at all redshifts; 

\smallskip 

A detailed analysis of the  two-point correlation functions $\xi_{\pm}$, as measured by KiDS21, shows that the \textit{Planck} \LCDM model can provide acceptable fits if the power-spectrum on non-linear scales is suppressed  via our phenomenological model of Eq. \ref{fig:Amod}, with a 
value $A_{\rm mod} \approx 0.69$. If this suppression is interpreted as being caused by baryonic feedback, then a comparison with numerical hydrodynamic simulations shows that strong baryonic feedback is required. The amplitude and spatial extent of the suppression that we require are well outside the ranges found in the BAHAMAS simulations and cannot be reproduced by baryonic feedback models
adopted in KiDS21 and \cite{Troester:2021extended} given their choices of priors.
 However, the physics of baryonic feedback is extremely complex and multi-scale and it may be premature to exclude models with strong baryonic feedback.  As mentioned in Sect.~\ref{subsec:TAGN}, 
combinations of cosmic shear with thermal and kinetic SZ measurements provide hints that baryonic feedback may be stronger than conventionally thought \citep{Amodeo:2021,schneider:2021, troester:2021}. In particular, kinetic SZ measurements offer the possibility of constraining the effects of baryonic feedback into the mildly non-linear regime. 

Recently some evidence for low values of $S_8$, though not at high statistical significance,  has come from cross-correlations of CMB lensing with photometric galaxy catalogues \citep{Marques:2020, Robertson:2021, Hang:2021, Krolewski:2021, Chang:2022, White:2021}, or with galaxy redshift surveys \citep{ChenWhite:2022}.
In some of these analyses, modifications to the non-linear 
power spectrum as  proposed in this paper may increase the inferred values of $S_8$. For others 
\citep[e.g.][]{ChenWhite:2022} the low values of $S_8$ are driven by cross-correlations at large angular scales
that are insensitive to the non-linear power spectrum but may be affected by selection biases. 
An analysis of galaxy-lensing cross-correlations shows that the preference for low values of $S_8$ comes from measurements at small scales \citep{Amon:2022}.

It is important to recognise the possibility  that a suppression of the  matter power spectrum on non-linear scales may be a consequence of new physics in the  dark sector. For example,  a suppression would arise if a fraction of the dark matter were in the form of a light axionic particle with a de Broglie wavelength of a few  Mpc \citep{Widrow:1993, Hu:2000, Hui:2017} though there are many other possibilities as discussed by \cite{Hooper:2022}. More detailed investigations of such models would be worthwhile, particularly if
it can be demonstrated definitively that baryonic feedback cannot solve the $S_8$ tension. In the longer term, it may be possible to differentiate between baryonic feedback and the physics of dark matter by measuring the redshift dependence of the matter power spectrum suppression \citep{Viel:2013} and using the high redshift reach of the quasar Lyman-$\alpha$ forest as discussed by
\cite{Hooper:2022}.

\section*{Acknowledgements}
We are particularly grateful to  Catherine Heymans and Hendrik Hildebrandt for a careful reading of a draft of this paper. We
 thank  Florian Buetler, Oliver Philcox, Licia Verde and Zvonomir Vlah for correspondence on redshift space distortions and Guido D'Amico, Leonardo Senatore and collaborators, for sharing their MCMC
 chains that we used in Fig. 1.  We indebted to Joe Zuntz for developing and maintaining the \textsc{cosmosSIS} framework \citep{Zuntz:2015}. 

AA receives support from a Kavli Fellowship at Cambridge University.  Based on observations made with ESO Telescopes at the La Silla Paranal Observatory under programme IDs 177.A-3016, 177.A-3017, 177.A-3018 and 179.A-2004, and on data products produced by the KiDS consortium. The KiDS production team acknowledges support from: Deutsche Forschungsgemeinschaft, ERC, NOVA and NWO-M grants; Target; the University of Padova, and the University Federico II (Naples).

\bibliographystyle{mnras} 
\bibliography{lensing}

\appendix

\section{COSEBIs}\label{sec:cosebis}

\begin{figure}
\hspace{-0.4truein}\includegraphics[width=110mm]{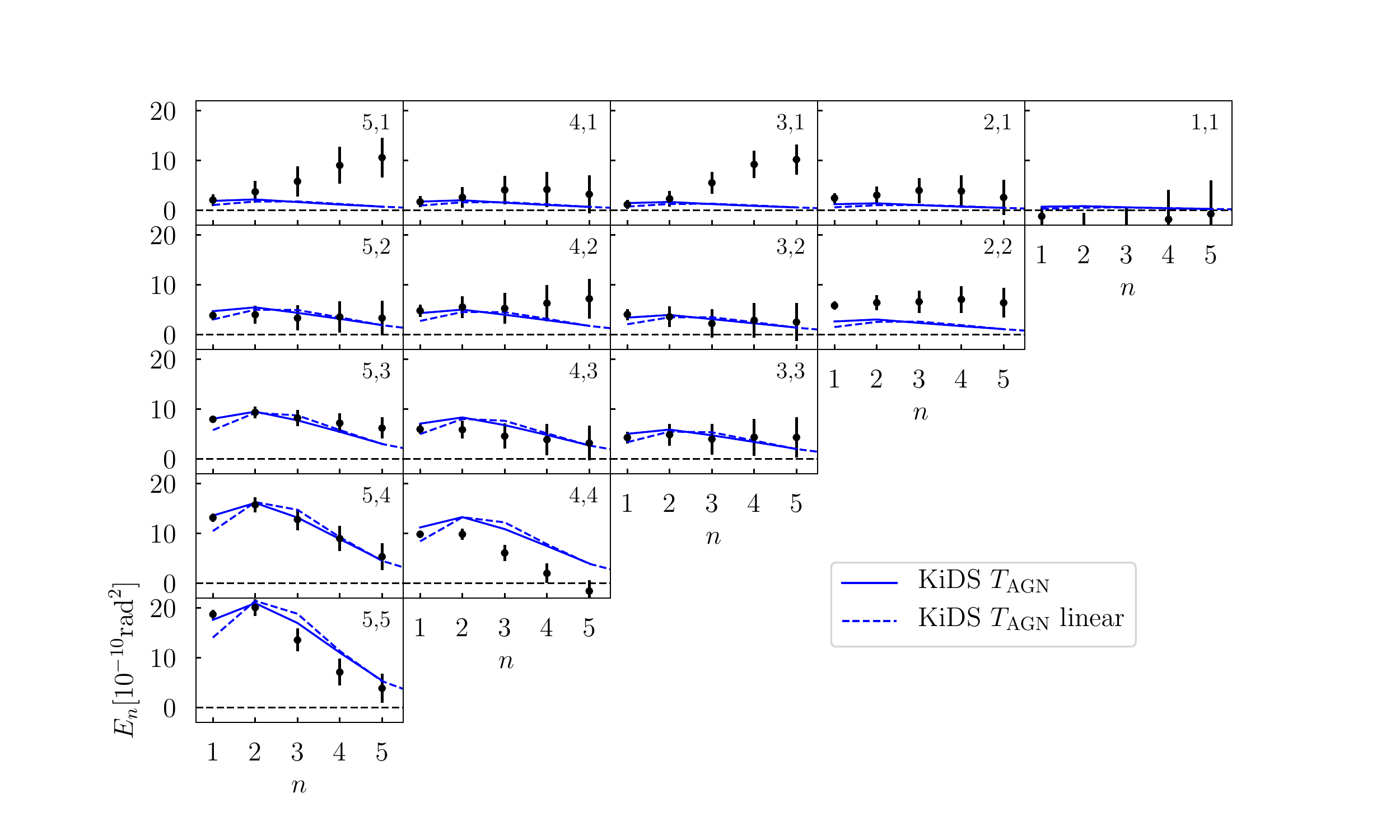}

	\caption{Measurements of KiDS21 COSEBI coefficients, $E_{\rm n}$, together with $1\sigma$ error bars. The numbers in each panel denote the photometric redshift  bins used in each cross-correlation as in Fig.~\ref{fig:xipm_kids} and a zero line is shown for reference (black dotted). The blue line is the best fit \LCDM theoretical prediction, including modelling of the non-linear matter power spectrum using {\textsc HMCode2020}	(very similar to the KiDS21 fit, which used {\textsc HMCode2016}), which includes a baryon feedback parameter $\Theta_{\rm AGN}$ as in our fiducial analysis. The dashed lines indicate the predictions based upon only the \textit{linear} matter power spectra. }
    \label{fig:COSEBI}
\end{figure}

The fiducial analysis in KiDS21 is based on the COSEBI statistic $E_n$.  The COSEBI statistics are constructed from linear combinations of $\xi_{\pm}$ to separate the signals from $E$- and $B$-mode lensing:
\begin{subequations}
\begin{eqnarray}
E_n  & = & {1 \over 2} \int_{\theta_{\rm min}}^{\theta_{\rm max}}
[T_{+n}(\theta) \xi_+(\theta) + T_{-n}(\theta) \xi_-(\theta)] \,, \\
B_n  & = & {1 \over 2} \int_{\theta_{\rm min}}^{\theta_{\rm max}} [T_{+n}(\theta) \xi_+(\theta) - T_{-n}(\theta) \xi_-(\theta)] \,,
\end{eqnarray}
\end{subequations}
where $n$ is an integer and the filter functions $T_{\rm n}$ depend on the choices of $\theta_{\rm min}$
and $\theta_{\rm max}$. Weak lensing should generate pure $E$ modes
and so the detection of a B-mode signal is a signature of systematics 
in the cosmic shear catalogues.

For KiDS-1000, the $B$-mode signal is 
consistent with zero. KiDS21 chose to analyze the first five $n$-modes of $E_n$, as plotted in Fig.~\ref{fig:COSEBI}. The weighting of wavenumbers in the statistic $E_n$ depends on the
range of $\theta_{\rm min}$ and $\theta_{\rm max}$. For the choices in
KiDS21, $0.5^\prime \le \theta \le 300^\prime$, the $E_n$ statistic is
gives lower weight to high wavenumbers in comparison to their  $\xi_{\pm}$ data vector and is therefore less sensitive to their  baryonic feedback model and other small-scale systematics \citep{Asgari:2020}.  

\begin{figure}
	\centering
	\vspace{0.3truein}
	\includegraphics[width=\columnwidth]{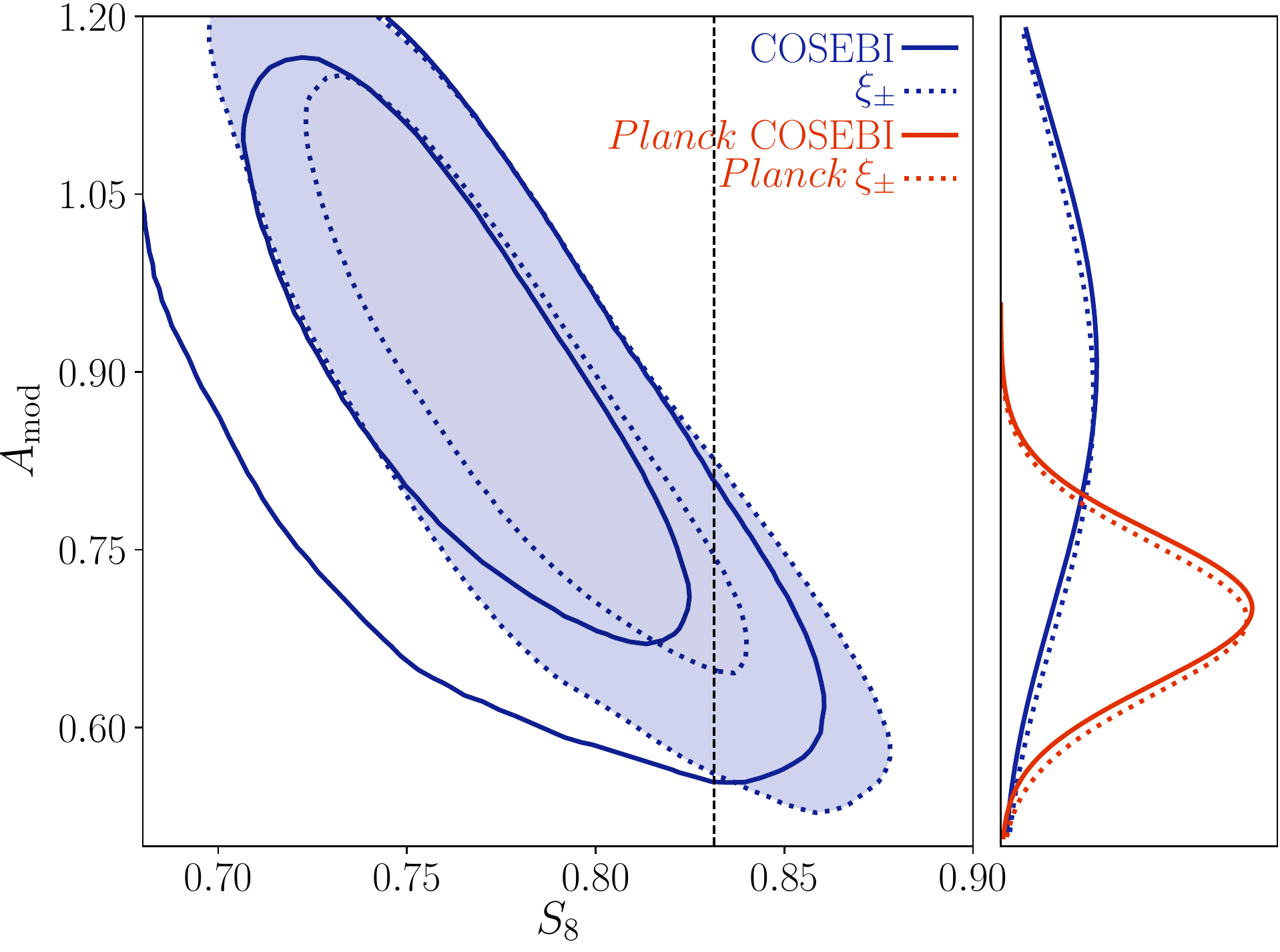} 
	\caption{Version of Fig.~\ref{fig:Amod} illustrating  the strong degeneracy between $S_8$ and  the phenomenological power spectrum suppression parameter $A_{\rm mod}$ but including results from the
	COSEBIS statistics. 
	The left hand panel shows the constraints from on $A_{\rm mod}$ and $S_8$ allowing cosmological and nuisance parameters to vary but with baryonic feedback set to zero. The solid blue contours  show the constraints using the COSEBIS, while the dotted blue contours
	show the results based on $\xi_{\pm}$  (which are the same as those 
	plotted in Fig.~\ref{fig:Amod}). 
	 The dashed line indicates the \Planck \LCDM best-fit value for the $S_8$ parameter.  The posteriors of $A_{\rm mod}$ are shown by the blue and dashed curves in the right hand panel. If the cosmological parameters are fixed to the  \Planck \LCDM best-fit values we find the posterior for $A_{\rm mod}$  shown by the red curves in the right hand panel.
	}
	\label{fig:AmodCOS}
\end{figure}

However, the $E_n$ statistic does retain sensitivity to non-linear scales and to baryonic physics, as is illustrated in Fig.~\ref{fig:AmodCOS} and Fig.~\ref{fig:COSEBI}. The solid line in Fig.~\ref{fig:COSEBI} shows the best fit to $E_n$ using  {\textsc HMCode2020}) with our fiducial prior on $\Theta_{\rm AGN}$. The dashed line shows the prediction based upon the linear power spectrum for this fit. One can see that the sensitivity to non-linearities (and therefore to baryonic feedback), for their choices of $\theta_{\rm min}$ and $\theta_{\rm max}$, is
restricted almost exclusively to the $E_1$ coefficients which carry much higher statistical weight in the likelihood than the coefficients with $n>1$. As a consequence, the $E_n$ statistic has very little shape discrimination. Figure~\ref{fig:AmodCOS} illustrates that the
COSEBIS and $\xi_{\pm}$ show a similarly
strong degeneracy between $A_{\rm mod}$  and $S_8$, though the COSEBI contours are considerably wider than those for $\xi_{\pm}$.

Another consequence of using COSEBIs is that they introduce a correlation between $S_8$ and $\Omega_{\rm m}$ that extends to low values of $S_8$ at high values of $\Omega_{\rm m}$ that are strongly disfavoured by \textit{Planck}. As discussed by KiDS21, the COSEBIs constraints on the parameter  combination $\Sigma_8 = \sigma_8 (\Omega_{\rm m}/0.3)^{0.54}$ are almost independent of $\Omega_{\rm m}$. In assessing consistency with \textit{Planck} \LCDM\ it is better to use the posterior distribution of $\Sigma_8$  rather than $S_8$.  Similar remarks apply to the \cite{troester:2021} analysis, which is based on the lensing power spectrum band powers.

\end{document}